\documentstyle[12pt,psfig]{article}

\newcommand{\AmS}{{\protect\the\textfont2
\renewcommand{\thesection}{\Roman{section}}
  A\kern-.1667em\lower.5ex\hbox{M}\kern-.125emS}}
 
\begin{document}
\rightline {DFTUZ 98/33}
\vskip 2. truecm
\centerline{\bf Rigorous arguments against current wisdoms in 
finite density QCD}
\vskip 2 truecm
\centerline { R. Aloisio$^{a,d}$, V.~Azcoiti$^b$, G. Di Carlo$^c$, 
A. Galante$^b$ and A.F. Grillo$^d$}
\vskip 1 truecm
\centerline {\it $^a$ Dipartimento di Fisica dell'Universit\`a 
dell'Aquila, L'Aquila 67100 (Italy).}
\vskip 0.15 truecm
\centerline {\it $^b$ Departamento de F\'\i sica Te\'orica, Facultad 
de Ciencias, Universidad de Zaragoza,}
\centerline {\it 50009 Zaragoza (Spain).}
\vskip 0.15 truecm
\centerline {\it $^c$ Istituto Nazionale di Fisica Nucleare, 
Laboratori Nazionali di Frascati,}
\centerline {\it P.O.B. 13 - Frascati 00044 (Italy). }
\vskip 0.15 truecm
\centerline {\it $^d$ Istituto Nazionale di Fisica Nucleare, 
Laboratori Nazionali del Gran Sasso,}
\centerline {\it Assergi (L'Aquila) 67010 (Italy). }
\vskip 3 truecm

\centerline {ABSTRACT}
\vskip 0.5truecm

\noindent
$QCD$ at finite chemical potential is analytically investigated in the 
region of large bare fermion masses. We show that, contrary to the general 
wisdom, the phase of the fermion determinant is irrelevant at zero 
temperature. However if the system is put at finite temperature, 
the contribution of the phase is finite.
We also discuss 
on the quenched approximation and suggest that the origin of the failure 
of this approximation in finite density $QCD$ could relay on the 
fundamental role that Pauli exclusion principle plays in this case.

\vfill\eject
\baselineskip=24pt
Numerical simulations of QCD at finite chemical potential $\mu$ are plagued 
by technical difficulties which, as well known, have delayed progress in 
this field. The non positivity of the determinant of the Dirac operator, 
together with indications coming from random matrix models, 
suggests that the phase of 
the determinant can not be neglected in the thermodynamical limit of the model.
The present situation is therefore rather pessimistic; it seems that there is 
no reliable hope to get important improvements in the knowledge of QCD 
at finite density from first principles. 

Having in mind these limitations, our aim here is to improve 
our knowledge of this field by analyzing QCD at finite $\mu$ in a limiting 
case which, even if far from the continuum limit of the model, 
allows one to do analytical manipulations forbidden in full QCD. The hope is 
that what is learned here can be of interest for the progress in this 
subject.

We are going to take the large bare fermion mass 
limit; the main simplification that follows is that all the 
temporal chains in the determinant of the Dirac operator decouple and the 
determinant can be written 
as a product of $V_s$ chains for every gauge configuration, 
$V_s$ being the number of space-like lattice 
points. 

We analyze QCD in this limit 
and will derive its phase diagram. We will show that, at $T=0$, 
the phase of the fermion determinant in the infinite volume limit does
not contribute to the free energy, a 
result against the general wisdom. The contrary happens at finite temperature.
Lastly, we show that, even at $T=0$, the Grand 
Canonical Partition Function approach leads to numerically uncorrect results,
not being able to reproduce the exact analytical features.

\section{Analytic results}

The Dirac-Kogut-Susskind operator of QCD at finite chemical potential can be 
written as 

\begin{equation}\label{1}
2 \Delta = 2mI + e^\mu G + e^{-\mu} G^\dagger + V
\end{equation}

\noindent
where $G$ ($G^+$) contains all forward (backward) temporal links and V all 
space-like links.

The determinant of $\Delta$ in the integration 
measure can be replaced, at large fermion masses $m$, by

\begin{equation}\label{2}
 \det \Delta = m^{3V_sL_t}\det \left( I + \frac{e^\mu}{2m} G \right)
\end{equation}

If the fugacity $e^\mu$ is much smaller than $2m$, the second factor of (2) 
can be replaced by 1 and the theory is independent of the chemical potential. 
Therefore, in order to get a non trivial $\mu$ dependence, we need to go 
to a region of large chemical potential in which the fugacity is of the 
order of $2m$ \cite{TOUS}.

Since all space-like links have disappeared in equation (2), the determinant 
of $\Delta$ factorizes as a product of $V_s$ determinants for the single 
temporal chains. A straightforward calculation allow us to write

\begin{equation}\label{3}
\det \Delta = e^{3V_sL_t\mu} \prod_{i=1}^{V_s} \det (c + L_i )
\end{equation}

\noindent
with $c=({2m\over{e^\mu}})^{L_t}$, $L_t$ is the lattice 
temporal extent and $L_i$ 
the SU(3) variable representing the forward straight Polyakov loop starting 
from the spatial site $i$ and circling once the lattice in the temporal 
direction. The determinants in (3) are gauge invariant quantities which can 
therefore be written as functions of the trace and the determinant of $L_i$. 
Since the gauge group is a unitary group, $\det(L_i)=1$ 
and therefore the only contributions depending on the gauge configuration 
will be functions of $Tr(L_i)$. In fact simple algebra allows to write

\begin{equation}\label{4}
\det (c + L_i ) = c^3 + c^2 Tr (L_i) + c Tr (L_i^*) + 1
\end{equation}
 
In the infinite gauge coupling limit, the integration over the gauge group 
is trivial since we get factorization \cite{LAT98}. 
The final result for the partition 
function at $\beta=0$ is

\begin{equation}\label{5}
{\cal Z} = 
V_G e^{3V_sL_t\mu}
\left( \left(\frac{2m}{e^\mu}\right)^{3L_t} +1 \right)^{V_s}
\end{equation}

\noindent
where $V_G$ is a constant irrelevant factor diverging exponentially 
with the lattice 
volume which accounts for the gauge group volume. Equation (5) 
gives for the free energy density $f={1\over{3V_sL_t}}\log{\cal Z}$

\begin{equation}\label{6}
f = \mu + 
\frac{1}{3L_t} \log \left( \left(\frac{2m}{e^\mu}\right)^{3L_t} +1 \right)
\end{equation}

The first contribution in (6) is an analytical function of $\mu$. The second 
contribution has, in the limit of infinite temporal lattice extent,  
a non analyticity at $\mu_c=\log(2m)$ which induces in the number density
a step jump, indication of a  saturation transition of first order at the  
value of $\mu_c$  previously given.

This is an expected result on physical grounds. In fact in the infinite 
fermion mass limit baryons are point-like particles, and  
pion exchange interaction vanishes, since pions are also very heavy. 
Therefore we are dealing with a system of very heavy free 
fermions (baryons) and by increasing the 
baryon density in such a system we expect an onset at 
$\mu_c={1\over3}m_b$, i.e., $\mu_c=\log(2m)$ since $3\log(2m)$ is the baryon 
mass at $\beta=0$ for large $m$ \cite{SACLAY}. 

Let us now discuss the relevance of the phase of the 
fermion determinant at $\beta=0$. The standard wisdom based on random matrix 
model results is that the phase of the fermion determinant plays a fundamental 
role in the thermodynamics of $QCD$ at finite baryon density \cite{RMT}
and that if 
the theory is simulated by replacing the determinant by its absolute value, 
one  neglects a contribution to the free energy density which could be 
fundamental in order to understand the critical behavior of this model. 
We are going to show now that, contrary to this wisdom, the phase of the 
determinant can be neglected in the large $m$ limit at $T=0$.

Equations (3)  and (4) imply that an upper bound for the absolute 
value of the fermion determinant is given by the determinant of the free 
gauge configuration. Therefore the mean value of the phase factor in the 
theory defined taking the absolute value of the determinant

\begin{equation}\label{7}
\left\langle e^{i\phi} \right\rangle_\| =
\frac{\int [dU] e^{-\beta S_G(U)}\det\Delta}
{\int [dU] e^{-\beta S_G(U)} | \det\Delta |}
\end{equation}

\noindent
is, at $\beta=0$, bounded from below by the ratio

\begin{equation}\label{8}
\left( \frac
{\left( \frac{2m}{e^\mu}\right)^{3L_t} + 1 }
{\left( \left( \frac{2m}{e^\mu}\right)^{L_t} + 1  \right)^3 }
\right)^{V_s}
\end{equation}

At zero temperature $(L_t=L, V_s=L^3)$, and letting $L\rightarrow\infty$,
it is straightforward to verify that the ratio (8) 
goes to 1 except at $\mu_c=\log(2m)$ 
(at $\mu=\mu_c$ the ratio
goes to zero but it is bounded from below by $(1/4)^{V_s}$).
Therefore the mean value of 
the cosine of the phase in the theory where the fermion determinant 
is replaced by its absolute value gives zero contribution.
 
At $T \neq 0$, i.e. taking the infinite $V_s$ limit by keeping 
fixed $L_t$, the lower bound 
(8) for the mean value of the phase factor (7) goes to zero exponentially 
with the spatial lattice volume $V_s$. This suggests that the 
phase will contribute in finite temperature $QCD$. 
In fact, it
 is easy to convince oneself that expression (7), at $\beta=0$, vanishes 
also exponentially with the lattice spatial volume at finite temperature
(see fig. 1). 
The contribution of the phase is therefore non zero (in the limit considered
here) in simulations of $QCD$ at finite temperature.

The free energy density at finite temperature (equation (6)) is an analytic 
function of the fermion mass and chemical potential. It develops a 
singularity only in the limit of zero temperature $(T={1\over{L_t}})$. 
Therefore $QCD$ at large $m$ and finite temperature does not show 
phase transition in the 
chemical potential but a crossover at $\mu=\log(2m)$ which becomes a true 
first order phase transition at $T=0$.

The standard way to define the theory at zero temperature is to consider
symmetric lattices.
However a more natural way to define the theory at $T=0$ 
is to take the limit of finite temperature $QCD$ when the physical 
temperature $T\rightarrow 0$. 
In other words, we should take first the infinite spatial 
volume limit and then the infinite temporal extent limit. 
We will show here that, as expected, physical results are independent of 
the procedure choosen.
The free energy density of the 
model can be written as the sum of two contributions $f=f_1+f_2$. The first 
contribution $f_1$ is the free energy density of the theory where the fermion 
determinant in the integration measure is replaced by its absolute value. 
The second contribution $f_2$, which comes from the phase of the fermion 
determinant, can be written as

\begin{equation}\label{9}
f_2 = {1\over{V_sL_t}}\log\left\langle e^{i\phi} \right\rangle_\|.
\end{equation}

\noindent
Since the mean value of the phase factor (7) is less or equal than 1, 
$f_2$ is  bounded from above by zero and from below by 

\begin{equation}\label{10}
{1\over{L_t}}\log{\left( \frac
{\left( \frac{2m}{e^\mu}\right)^{3L_t} + 1 }
{\left( \left( \frac{2m}{e^\mu}\right)^{L_t} + 1  \right)^3 }
\right)}
\end{equation}

When $L_t$ goes to infinity, expression (10) goes to zero for all the
values of $\mu$ and therefore the 
only contribution to the free energy density which survives in the zero 
temperature limit is $f_1$.
Again, we conclude that zero temperature QCD in the strong coupling 
limit at finite chemical potential 
and for large fermion masses is well described by the theory obtained by 
replacing the fermion determinant by its absolute value.

These results are not surprising as follows from the fact that at $\beta=0$ 
and for large $m$ the system factorizes as a product of $V_s$ noninteracting 
$0+1$ dimensional $QCD's$ and from the relevance (irrelevance) of the phase 
of the fermion determinant in $0+1$ QCD at finite (zero) "temperature"
\cite{LAT97}. 
More surprising maybe is that, as we will see in the following, 
some of these results do not change when we put a finite gauge coupling. 

The inclusion of a non trivial pure gauge Boltzmann factor in the integration 
measure of the partition function breaks the factorization property. The 
effect of a finite gauge coupling is to induce correlations between the 
different temporal chains of the determinant of the Dirac operator. The 
partition function is given by

\begin{equation}\label{11}
{\cal Z} = \int [dU] e^{-\beta S_G(U)} \prod_{i=1}^{V_s}
(c^3 + 1 +  c Tr (L_i^*) + c^2 Tr (L_i)  )
\end{equation}

\noindent
and can be written as 

\begin{equation}
{\cal Z}(\beta,\mu) = {\cal Z}_{pg}\cdot {\cal Z}(\beta=0,\mu)\cdot 
R(\beta,\mu)
\end{equation}

\noindent
where ${\cal Z}_{pg}$ is the pure gauge partition function, 
${\cal Z}(\beta=0,\mu)$ the strong coupling partition function 
(equation (5)) and $R(\beta,\mu)$ is given by

\begin{equation}\label{12}
R(\beta,\mu) = \frac
{\int [dU] e^{-\beta S_G(U)} \prod_{i=1}^{V_s} \left(
1 + \frac{c Tr (L_i) +  c^2 Tr (L_i^*)}{c^3 + 1} \right)}
{\int [dU] e^{-\beta S_G(U)}}
\end{equation}

In the zero temperature limit ($L_t=L, L_s=L^{3}, L\rightarrow\infty$) the 
productory in the numerator of (13) goes to 1 independently of the gauge 
configuration. In fact each single factor has an absolute value equal to 
1 up to corrections which vanish exponentially with the lattice size $L$ 
and a phase which vanishes also exponentially with $L$. Since the total number 
of factors is $L^3$, the productory goes to 1 and therefore $R=1$ in the 
zero temperature limit. 

The contribution of $R$ to the free energy 
density vanishes therefore in the infinite volume limit at zero temperature. 
In such a case, the free energy density is the sum of the free energy density 
of the pure gauge $SU(3)$ theory plus the free energy density of the model at 
$\beta=0$ (equation (6)). The first order phase transition found at $\beta=0$ 
is also present at any $\beta$ and its location and properties do not 
depend on $\beta$ since all $\beta$ dependence in the partition function 
factorizes in the pure gauge contribution. 
Again at finite gauge coupling 
the phase of the fermion determinant is irrelevant at zero temperature.

At finite temperature and finite gauge coupling the first order phase 
transition induced by the contribution (6) to the free energy density at 
zero temperature disappears and becomes a crossover. 
Furthermore expression (13) gives also a non vanishing contribution 
to the free energy density if $L_t$ is finite.  

The common physical interpretation for
the theory with the absolute value of the fermion determinant 
is that it possesses 
quarks in the {\bf 3} and {\bf 3}$^*$ representations of SU(3),
having baryonic states made up of two quarks which would give account for the
physical differences respect to real QCD. We have proven analytically (at
$\beta=0$) that the relation between modulus and real QCD is 
temperature dependent, $i.e.$ they are different only at $T \ne 0$, 
a feature that does not support the above interpretation.

\section{Numerical results}

From the point of view of simulations, work has been done by 
several groups mainly to develop numerical algorithms capable to overcome
the non positivity of the fermionic determinant. 
The most promising of these algorithms \cite{BAR}, \cite{NOI1}
are based on the GCPF formalism and try to calculate extensive quantities 
(the canonical partition functions at fixed baryon number).
Usually they measure quantities that, with actual statistics, do not 
converge. 

In a previous paper \cite{NOI2} we have given arguments to conclude that, 
if the phase is relevant, a statistics exponentially increasing 
with the system volume 
is necessary to appreciate its contribution to the observables (see also
\cite{BAR2} ).
What happens if we consider a case where the phase is not relevant
($i.e.$ the large mass limit of QCD at zero temperature, as discussed
in the previous section)?

To answer this question we have reformulated the GCPF formalism by
writing the partition function as a polynomial in 
$c$ and studied the convergence properties of the coefficients at $\beta=0$
using an ensemble of (several thousands) random configurations.
This has been done as in standard numerical simulations ({\it i.e.} 
without using the factorization property) for lattices $4^4$ (fig. 2a), 
$4^3\times 20$ (fig. 2b), $10^3\times 4$ (fig. 2c) \cite{LAT98} and the 
results compared with the analytical predictions (\ref{5}) (solid lines 
in the figures).

From these plots we can see that, unless we consider a large lattice temporal
extension, our averaged coefficients in the infinite coupling limit still
suffer from sign ambiguities $i.e.$ not all of them are positive.
For large $L_t$ the {\it sign problem} 
tends to disappear because the determinant of the one dimensional system 
(\ref{4}) becomes an almost real quantity for each gauge configuration and 
obviously the same happens to the determinant of the Dirac operator
(\ref{3}) in the four dimensional lattice. 
It is also interesting to note that the sign of the averaged
coefficients is very stable and a different set of random configurations
produces almost the same shape. 

However, the sign of the determinant is not the only problem: in fact,
as one can read  from fig. 2, even considering the modulus of
the averaged coefficients we do not get the correct result. 
We used the same configurations to calculate the average of the modulus of
the coefficients. We expect this quantity to be larger 
than the analityc results reported in fig. 2.
The data, however, contrast with this scenario: 
the averages of the modulus are always smaller (on a logarithmic scale) 
than the analityc results from formula (\ref{5}).
In fact these averages are indistinguishable from the absolute values of 
the numerical results reported in fig. 2.

In conclusion, even if the phase of the fermion
determinant is irrelevant in QCD at finite density ($T=0$ and heavy quarks) 
the numerical evaluation of the Grand Canonical Partition Function 
still suffers from sampling problems.

A last interesting feature which can be discussed on the light of our results 
concerns the validity of the quenched approximation in finite density $QCD$. 
An important amount of experience in this field \cite{KOGUT} suggests that 
contrary to what happens in $QCD$ at finite and zero temperature, the quenched 
approximation does not give correct results in $QCD$ at finite chemical 
potential. Even if 
the zero flavour limit of the theory with the absolute value of the 
fermion determinant and of actual $QCD$ are the same (quenched approximation), 
the failure of this approximation has been assigned in the past \cite{RMT} 
to the fact that it corresponds to the zero flavour limit of the theory 
with $n$ quarks in the fundamental and $n$ quarks in the complex 
representation of 
the gauge group. In fig. 3 we have plotted the number density at $\beta=0$ and 
for heavy quarks in three interesting cases: actual $QCD$, the theory 
with the absolute value of the fermion determinant and quenched $QCD$. 
It is obvious that the quenched approximation produces results far from those 
of actual $QCD$ but also far from those of $QCD$ with the modulus of the 
determinant of the Dirac operator. The former results are furthermore very 
near to those of actual $QCD$. In other words, even if the phase is relevant 
at finite temperature, its contribution to the number density is almost 
negligible. 

It seems unplausible on the light of these results to assign the failure of 
the quenched approximation to the feature previously discussed \cite{RMT}. 
It seems more natural to speculate that it fails because does not 
incorporate correctly in the path integral the Fermi-Dirac statistics and 
we do expect that Pauli exclusion principle play, by far, a more relevant 
role in finite density $QCD$ than in finite temperature $QCD$.

\vskip 0.3truecm
\noindent
{\bf Acknowledgements}
\vskip 0.3truecm

This work has been partially supported by CICYT and INFN.

\newpage

\begin{figure}[!t]\
\psrotatefirst
\psfig{figure=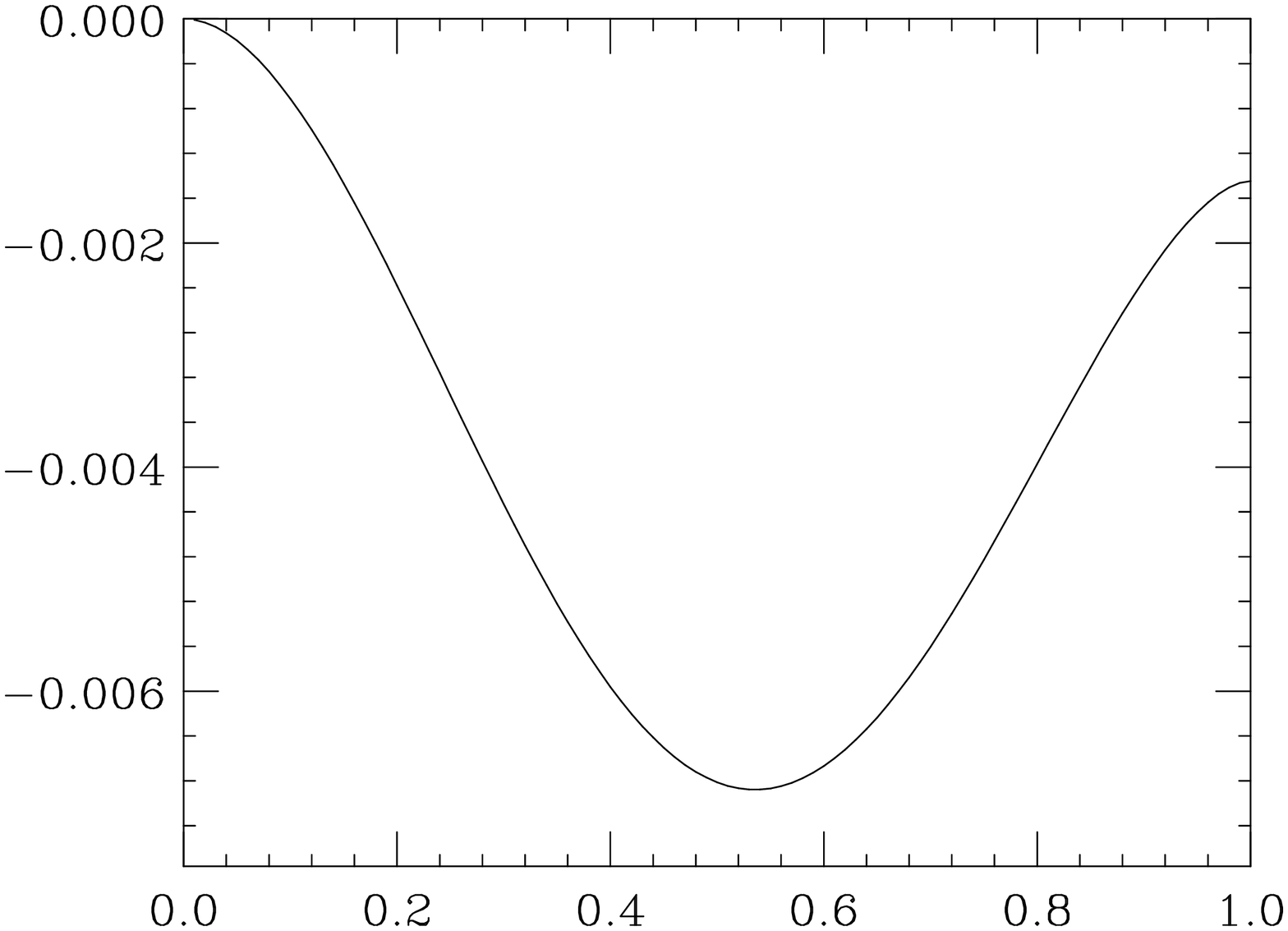,angle=90,width=400pt}
\caption{Inverse temperature in lattice units times the contribution 
to the free energy density coming from the phase. This quantity is evaluated 
in the strong coupling limit and plotted as a function of $c^{-1}$; it is
invariant under the transformation $c\to c^{-1}$.}
\label{fig1}
\end{figure}

\newpage

\begin{figure}[!t]\
\hskip 2truecm
\begin{minipage}[t]{75mm}
\psrotatefirst
\psfig{figure=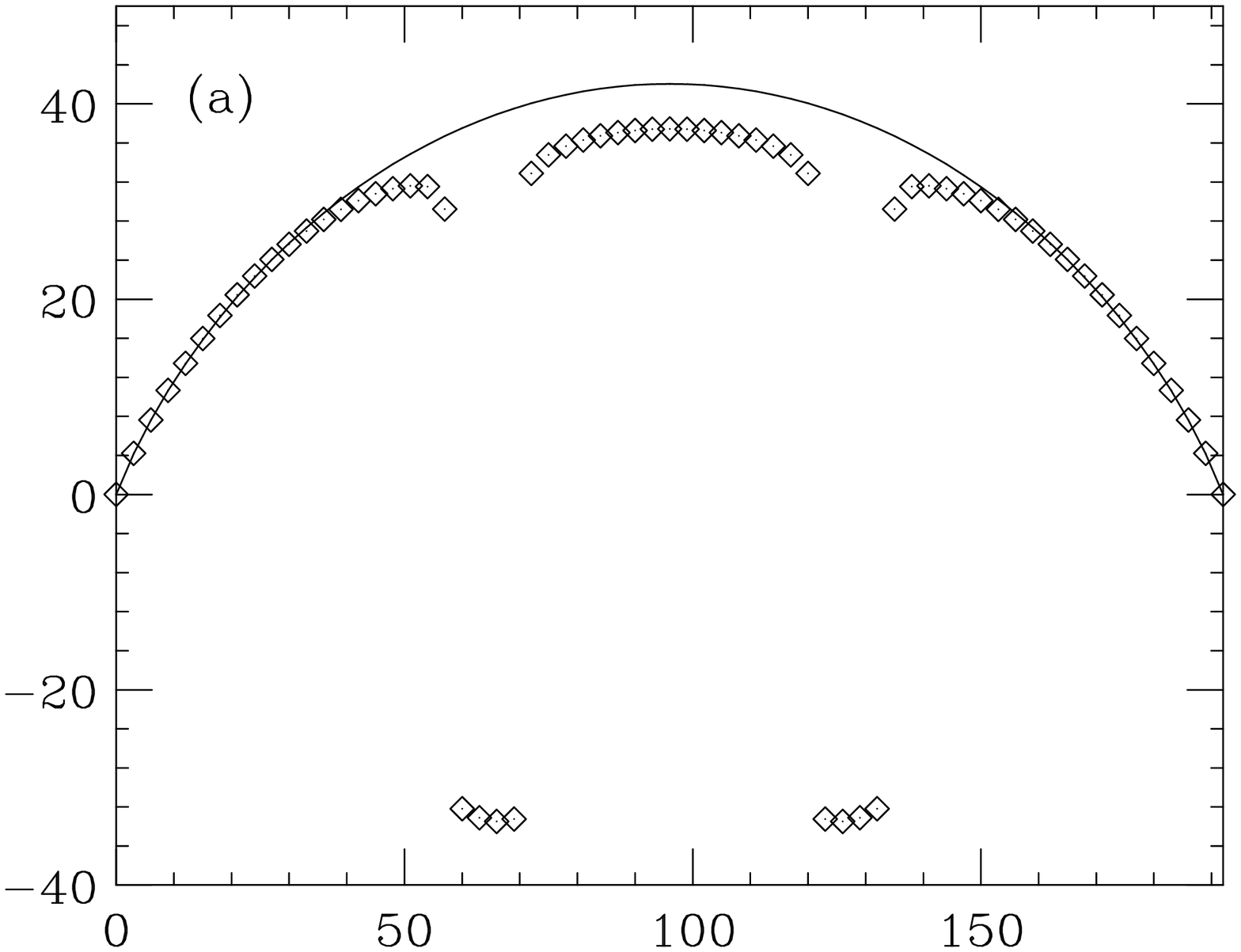,angle=90,width=220pt}
\end{minipage}
\hspace{\fill}
\vskip 0.1truecm
\hskip 2.3truecm
\begin{minipage}[t]{75mm}
\psrotatefirst
\psfig{figure=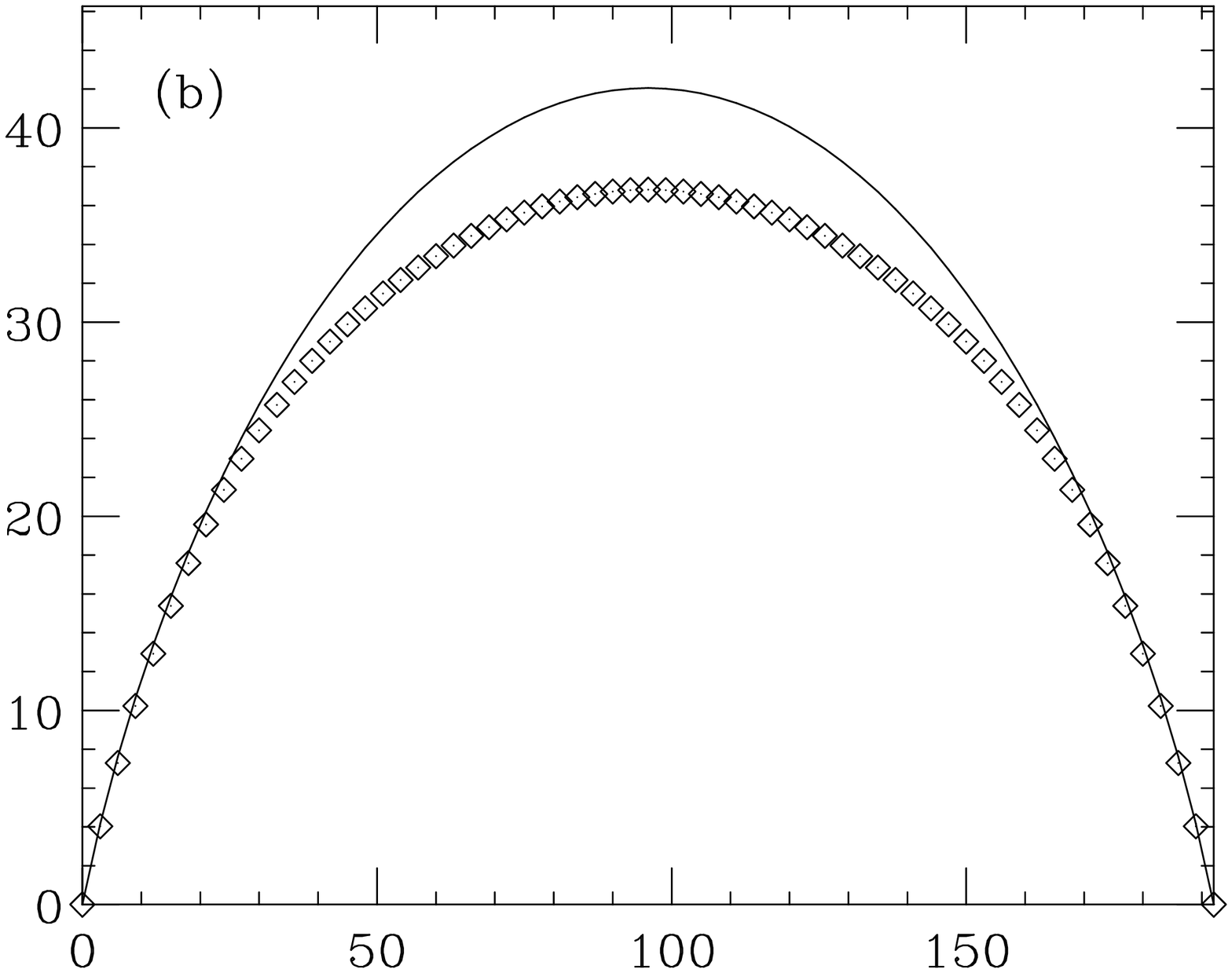,angle=90,width=215pt}
\end{minipage}
\hspace{\fill}
\vskip 0.1 truecm
\hskip 2truecm
\begin{minipage}[t]{75mm}
\psrotatefirst
\psfig{figure=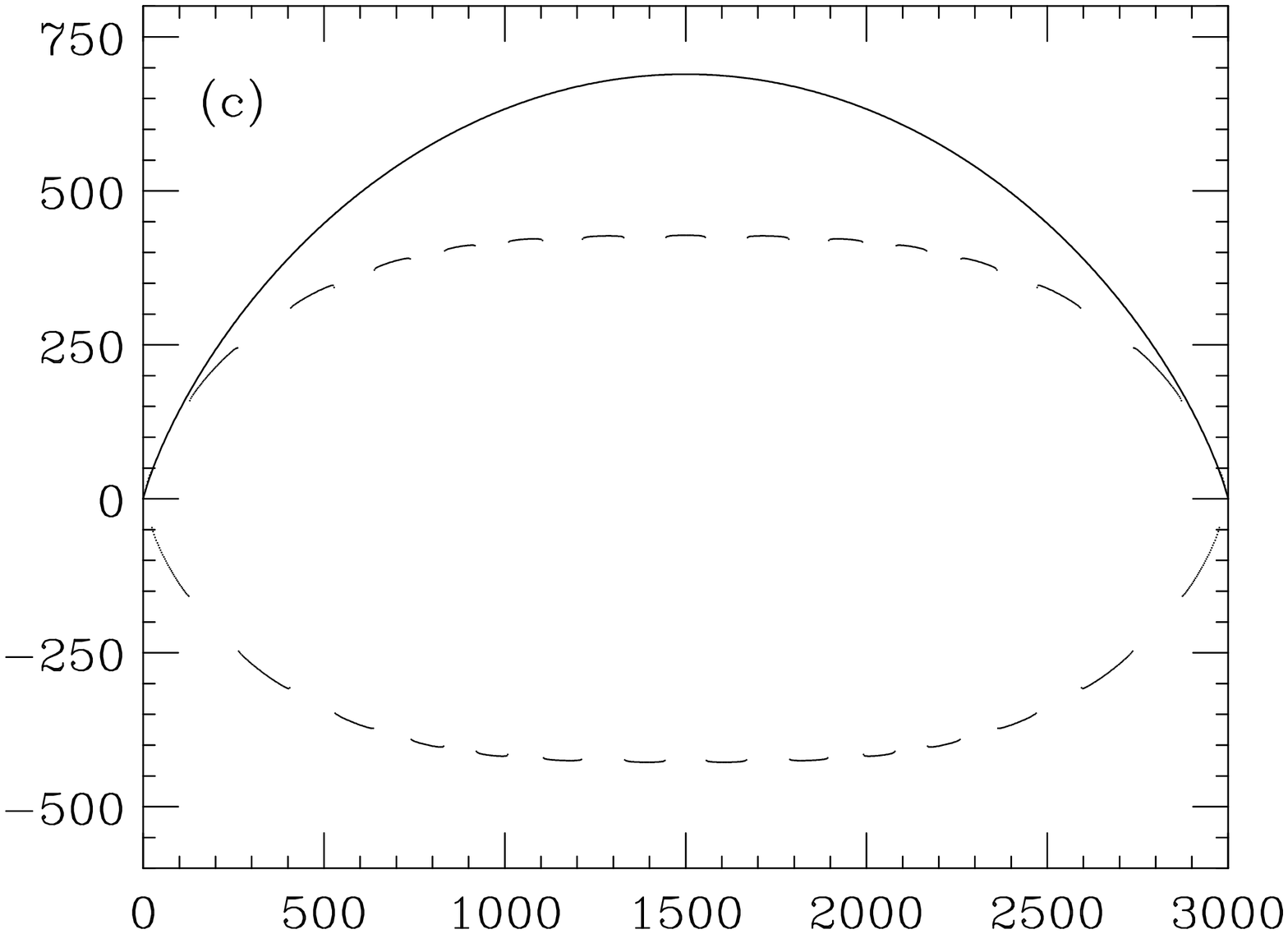,angle=90,width=230pt}
\end{minipage}
\caption{Logarithm of the modulus of the GCPF coefficients times the sign of
the coefficients. $V=4^4\times 4$ (a), $4^3\times 20$ (b), $10^3\times 4$ (c);
the uppermost continuous curve are the analytic results.}
\label{fig2}
\end{figure}

\newpage

\begin{figure}[!t]\
\psrotatefirst
\psfig{figure=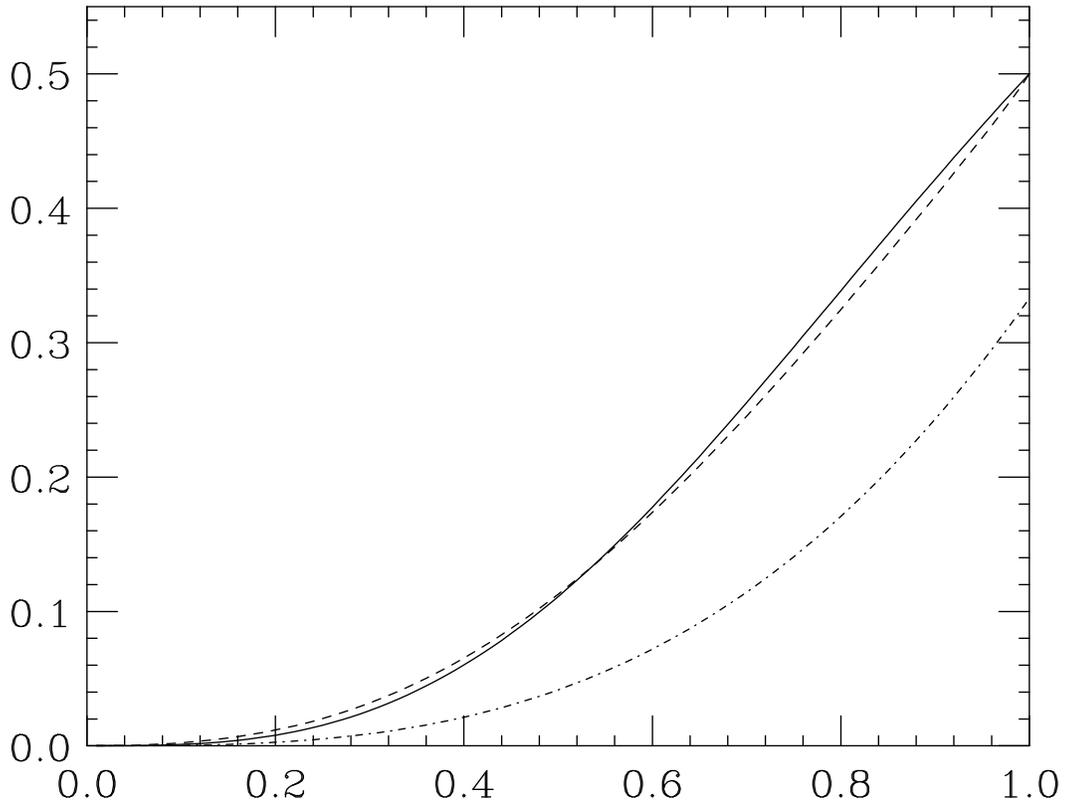,angle=90,width=400pt}
\caption{Baryonic density $n$ as a function of $c^{-1}$ for 
the complete theory (continuos line), the theory defined using the modulus 
(dashed line) and the quenched theory (dot-dashed line). 
All the results reported in this figure have been obtained analitically 
(we can use the relation $n(c)=1-n(1/c)$ to reconstruct the whole curves).}
\label{fig3}
\end{figure}

\end{document}